\documentclass[twocolumn,english,aps,prl,showpacs,superscriptaddress]{revtex4-1}
\usepackage[T1]{fontenc}
\usepackage[latin9]{inputenc}
\usepackage{color}
\usepackage{babel}
\usepackage{amsmath}
\usepackage{amssymb}
\usepackage{graphicx}
\usepackage[unicode=true,
 bookmarks=true,bookmarksnumbered=false,bookmarksopen=false,
 breaklinks=false,pdfborder={0 0 0},pdfborderstyle={},backref=false,colorlinks=true]
 {hyperref}
\begin{document}

\title{Comparison of quantum kinetic theory and time-dependent Dirac equation
approaches in vacuum pair production and the bound states resonance
enhanced mechanism}

\author{Q. Wang}

\affiliation{Institute of Applied Physics and Computational Mathematics, Beijing
100088, China}

\author{L. B. Fu}
\email{lbfu@iapcm.ac.cn}

\affiliation{Graduate School, China Academy of Engineering Physics, Beijing 100193,
China}

\affiliation{HEDPS, Center for Applied Physics and Technology, Peking University,
Beijing 100871, China~\\
 and CICIFSA MoE College of Engineering, Peking University, Beijing
100871, China}

\author{J. Liu}

\affiliation{Institute of Applied Physics and Computational Mathematics, Beijing
100088, China}

\affiliation{HEDPS, Center for Applied Physics and Technology, Peking University,
Beijing 100871, China~\\
 and CICIFSA MoE College of Engineering, Peking University, Beijing
100871, China}
\begin{abstract}
A remarkable quantitative agreement is found between the non-Markovian
quantum kinetic approach and the time-dependent Dirac equation approach
for a large region of Keldysh parameter, in the investigation of electron-positron
pair production in the electric fields which is spatially homogeneous
and envelope pulse shaped. If a sub-critical bound potential is immersed
in this background field, the TDDE results show that the creation
probability will be enhanced by the bound states resonance by two
orders of magnitude. We also establish a computing resources greatly
saved TDDE formalism for spatially homogeneous field.
\end{abstract}


\maketitle

\textbf{Introduction.} In the presence of a very strong electric field
the quantum electrodynamics (QED) vacuum may break down and decay
via the production of electron-positron pairs. This effect was first
discussed by Sauter\citep{Sauter} and computed by Schwinger for the
spatially homogeneous and static electric fields\citep{Schwinger}.
Since then various theoretical techniques were developed to deal with
more complicate field configurations\citep{manymethods}. 

In recent years, the quantum kinetic theory (QKT) based on Vlasov
equations including a source term was established to resolve the dynamics
of the production process which is in general non-equilibrium and
time-dependent \citep{VlasovEq_Markovlimit,QKT}. QKT is rigorously
derived form QED by canonical quantization of the Dirac field and
subsequently Bogoliubov transformation to a quasi-particle representation.
It is exact on the mean-field level. In the sub-critical field strength
regime, the collision \citep{neglectc,neglectcb} and back reaction\citep{neglectcb,backreaction,neglectb}
terms can be neglected safely. It is easily to implement numerical
calculation, and can provide not only the pair production rate but
also the momentum distribution information. These advantages make
it a powerful tool to investigate pair creation in complicated temporal
field configurations\citep{QKTworks,effectivemass}. QKT has been
proved to be equivalent to the DHW formalism in the case of linearly
polarized electric fields\citep{EquivalenceDHW}, and to the scattering
approach\citep{EquivalenceS}. One the other hand, the disadvantage
of QKT is obvious that its application is restricted to spatial homogeneous
field, thus to one dimension.

In a realistic experiment, the spatial variation of the field should
be taken into account. In current literature, the electron-positron
pair number can be obtained by the time-dependent Dirac equation approach
(TDDE). Pair production in laser fields oscillating in space and time
is investigated by propagating an initially negative-energy Gaussian
wave packet in the spatial- and temporal- dependent fields and projecting
it onto positive energy states after the fields has been turned off
\citep{MR09}. This approach based on one-particle time-dependent
Dirac equation has rigorous foundation in the well-established revised
version of Furry\textquoteright s formulation of QED in external fields
with unstable vacuum\citep{foundation1,foundation2,foundation3}.
Another approach is the called numerical quantum field theory method,
where the pair number is obtained by propagating all the negative
energy states in the time dependent Dirac Hamiltonian and projecting
them over all positive energy states \citep{TDDESu}. In the following,
the abbreviation 'TDDE' denotes the later. In contrast to QKT, in
TDDE approach the spatial variation of the field can be taken into
account, and the information of spatial and momentum distribution
of both electrons and positrons can be computed easily as well. In
principle, if applied in homogeneous fields, the TDDE approach must
be equivalent to QKT. However, surprisingly, few works have been done
to examine this equivalence. This will be the first goal of this paper.

In a recent work\citep{enhanced}, it is suggested that if additional
binding potentials (like that of a bare nucleus) are immersed in the
constant electric field region, the corresponding bound state can
enhance the pair creation. Of course, similar to the charge resonance
enhanced ionization of molecular physics\citep{resonance_enhanced_ioniz1,resonance_enhanced_ioniz2,resonance_enhanced_ioniz3},
bound states (for example, supported by a super strong nuclear Coulomb
field characteristic of two colliding high-Z ions) play an important
role in the pair creation process. In the work\citep{STang}, where
bound states located in the gap supported by the well potential are
exposed in an oscillating electric field, an simple matchup between
bound states and momentum distribution was found. The second goal
of the letter is, in a real binding potential, examining the bound
states enhanced pair creation using the TDDE approach.

In the following, we will first briefly review the two approaches,
and establish a formalism for TDDE in spatial homogeneous case. The
natural units are used, that $\hbar=e=c=1$ with other quantities
scaled by $m$. Then, for a homogeneous time-dependent electric field
pulse, we compare and discuss the numerical results. Finally we show
the bound states resonance enhanced pair creation. 

\textbf{Quantum Kinetic Theory approach (QKT).} 

Assuming the vector potential is $\mathbf{A}=\left(0,0,A_{z}\left(t\right)\right)$,
since momentum ${\bf k}(k_{\perp},k_{z})$ is a good quantum number,
the field operator $\Phi\left(\mathbf{x},t\right)$ is decomposed
as $\Phi(\mathbf{x},t)=\int d{\bf k}e^{i{\bf k\cdot x}}\left(u_{{\bf k}}(t)\tilde{a}_{{\bf k}}+u_{{\bf k}}^{*}(t)\widetilde{b}_{-{\bf k}}^{\dagger}\right)$,
where $\tilde{a}_{{\bf k}}$ and $\tilde{b}_{{\bf -k}}^{\dagger}$
correspond to the annihilation (creation) operators for the particle
and the antiparticle respectively with the momentum $|\mathbf{k}|$.
The transformation between time dependent operators ($a_{{\bf k}}\left(t\right),b_{{\bf k}}\left(t\right)$)
and time independent operators ($\tilde{a}_{{\bf k}},\tilde{b}_{{\bf k}}$)
is expressed in terms of the Bogoliubov coefficients $\alpha_{{\bf k}}(t)$
and $\beta_{{\bf k}}(t)$ ,

\begin{eqnarray}
a_{{\bf k}}\left(t\right) & = & \alpha_{{\bf k}}(t)\tilde{a}_{{\bf k}}-\beta_{{\bf k}}^{*}(t)\tilde{b}_{-{\bf k}}^{\dagger},\label{eq:bgQKT1}\\
b_{-{\bf k}}^{\dagger}\left(t\right) & = & \beta_{{\bf k}}(t)\tilde{a}_{{\bf k}}+\alpha_{{\bf k}}^{*}(t)\tilde{b}_{-{\bf k}}^{\dagger},\label{eq:bgQKT2}
\end{eqnarray}
with $|\alpha_{{\bf k}}(t)|^{2}+|\beta_{{\bf k}}(t)|^{2}=1.$ By making
an adiabatic ansatz for $u_{{\bf k}}(t)$\citep{EquivalenceS}, the
Dirac equation which $\Phi(\mathbf{x},t)$ obeys reduces to an oscillator
equation $\ddot{u}_{{\bf k}}(t)+\left(\omega^{2}\left({\bf k},t\right)+i\dot{P}_{z}\left(t\right)\right)u_{{\bf k}}(t)=0$,
where $\omega^{2}\left({\bf k},t\right)=m^{2}+k_{\bot}^{2}+P_{z}^{2}$
, $P_{z}\left(t\right)=k_{z}+A_{z}\left(t\right)$. The number of
pairs for each ${\bf k}$ is defined as $N\left({\bf k},t\right)=|\beta_{{\bf k}}(t)|^{2}$.
The adiabatic ansatz gives the creation rate as the source term of
the quantum Vlasov equation, $\dot{N}\left({\bf k},t\right)=W\left({\bf k},t\right)\int_{-\infty}^{t}W\left({\bf k},t\right)\left(1-N\left({\bf k},t'\right)\right)\cos\left(2\int_{t}^{t'}\omega({\bf k},\tau)d\tau\right)dt',$
where $W\left({\bf k},t\right)=\frac{eE_{z}\left(t\right)\epsilon_{\bot}\left(t\right)}{\omega^{2}({\bf k},t)}$,
$\epsilon_{\bot}^{2}\left(t\right)=k_{\bot}^{2}+m^{2}$, $E\left(t\right)$
is the electric field along $z$ direction $E_{z}\left(t\right)=-\dot{A_{z}}\left(t\right)$,
and $P_{z}\left(t\right)$ denotes the time-dependent kinetic momentum.
For computational efficiency, by introducing two auxiliary functions
$G\left({\bf k},t\right)$ and $H\left({\bf k},t\right)$, the quantum
Vlasov equation can be re-expressed as first-order differential equations\citep{backreaction},

\begin{eqnarray}
\dot{N}\left({\bf k},t\right) & = & W\left({\bf k},t\right)G\left({\bf k},t\right),\\
\dot{G}\left({\bf k},t\right) & = & W\left({\bf k},t\right)\left[1-N\left({\bf k},t\right)\right]-2\omega\left({\bf k},t\right)H\left({\bf k},t\right),\\
\dot{H}\left({\bf k},t\right) & = & 2\omega\left({\bf k},t\right)G\left({\bf k},t\right),
\end{eqnarray}
with initial conditions $N\left({\bf k},-\infty\right)=G\left({\bf k},-\infty\right)=H\left({\bf k},-\infty\right)=0$.
Here $N$ accounts for both spin directions. These equations can be
solved easily using Runge-Kutta methods. Throughout the article, we
perform all simulations in one dimension, $k_{\bot}=0$. Then the
particle yield per Compton wavelength ($\lambda_{C}$) reads $N=\int dk_{z}/\left(2\pi\right)N\left(k_{z},+\infty\right)$. 

\textbf{Time-dependent Dirac Equation (TDDE).} 

Here the field operator $\ensuremath{\hat{\Psi}\left(\vec{\boldsymbol{x}},t\right)}$
is also expressed in terms of the electron annihilation $\hat{b}$
and positron creation $\hat{d}^{\dagger}$ operators,$\hat{\Psi}\left(\vec{\boldsymbol{x}},t\right)=\sum_{p}\hat{b}_{p}\varphi_{p}\left(\vec{\boldsymbol{x}},t\right)+\sum_{n}\hat{d}_{n}^{\dagger}\varphi_{n}\left(\vec{\boldsymbol{x}},t\right)$,
and the relations between the time dependent and time independent
operator are,

\begin{eqnarray}
\hat{b}_{p}\left(t\right) & = & \sum_{p'}\hat{b}_{p'}U_{pp'}\left(t\right)+\sum_{n'}\hat{d}_{n'}^{\dagger}U_{pn'}\left(t\right),\label{eq:bgTDDE1}\\
\hat{d}_{n}^{\dagger}\left(t\right) & = & \sum_{p'}\hat{b}_{p'}U_{np'}\left(t\right)+\sum_{n'}\hat{d}_{n'}^{\dagger}U_{nn'}\left(t\right).\label{eq:bgTDDE2}
\end{eqnarray}
Particle information can be obtained from the field operator, e.g.,
the spacial distribution of electrons created from the vacuum ( defined
as $\hat{b}_{p}\left\Vert vac\right\rangle =0$, $\hat{d}_{n}\left\Vert vac\right\rangle =0$)
is obtained from the positive part of the field operator, $N^{el.}\left(\vec{\boldsymbol{x}},t\right)=\left\langle vac\right\Vert \hat{\Psi}^{(+)\dagger}\left(x,t\right)\hat{\Psi}^{(+)}\left(x,t\right)\left\Vert vac\right\rangle $.
The pair number reads, $N\left(t\right)=\sum_{pn}\left|U_{pn}\left(t\right)\right|^{2}$.
The density and momentum distribution of electrons are $N^{el.}\left(\vec{\boldsymbol{x}},t\right)=\sum_{n}\left|\sum_{p}U_{pn}\left(t\right)\varphi_{p}\left(\vec{\boldsymbol{x}}\right)\right|^{2},$
$N^{el.}\left(p,t\right)=\sum_{n}\left|U_{pn}\left(t\right)\right|^{2}.$
These quantities of positrons can also be computed, $N^{po.}\left(\vec{\boldsymbol{x}},t\right)=\sum_{p}\left|\sum_{n}U_{pn}\left(t\right)\varphi_{n}\left(\vec{\boldsymbol{x}}\right)\right|^{2}$,
$N^{po.}\left(n,t\right)=\sum_{p}\left|U_{pn}\left(t\right)\right|^{2}$.
Here the transition probability $U_{pn}\left(t\right)=\int d\vec{\boldsymbol{x}}\varphi_{p}^{*}\left(\vec{\boldsymbol{x}}\right)\varphi_{n}\left(\vec{\boldsymbol{x}},t\right)$
is computed by propagating every initial negative energy eigen-states
by spatial- and temporal- dependent Dirac Hamiltonian, and then projecting
them over all positive energy eigen-states. $\sum_{p(n)}$ denote
the summation over all states with positive (negative) energy. $\varphi_{p\left(n\right)}\left(\vec{\boldsymbol{x}}\right)$
are the positive (negative) energy eigen-states of the field-free
Dirac Hamiltonian, $\varphi_{p\left(n\right)}\left(\vec{\boldsymbol{x}},t\right)$
are solutions of the time dependent Dirac equation with potential
taking into account, $\varphi_{p\left(n\right)}\left(\vec{\boldsymbol{x}},t\right)=\hat{U}\left(t,-\infty\right)\varphi_{p\left(n\right)}\left(\vec{\boldsymbol{x}}\right)$,
and can be obtained using the numerical split operator technique\citep{split}.
The time-evolution operator is defined as $\hat{U}\left(t_{2},t_{1}\right)=\hat{T}\mathrm{exp}\left(-i\int_{t_{1}}^{t_{2}}dt'\hat{H}\left(t'\right)\right)$,
where $\hat{T}$ denotes the Dyson time ordering operator. The Dirac
Hamiltonian is $H=\boldsymbol{\alpha}\cdot\boldsymbol{p}+m\beta-A_{0}+\boldsymbol{\alpha}\cdot\vec{A}$,
where $A_{0}$ and $\vec{A}$ denote scalar and vector potentials,
$\vec{E}=-\nabla A_{0}-\dot{\vec{A}}$. Each temporal step reads $\varphi\left(\vec{\boldsymbol{x}},t+dt\right)\approx\exp\left(-i\frac{dt}{2}H_{\partial}\right)\exp\left(-idtH_{\vec{\boldsymbol{x}}}\right)\exp\left(-i\frac{dt}{2}H_{\partial}\right)\varphi\left(\vec{\boldsymbol{x}},t\right)+O\left(dt^{3}\right)$,
where $H_{\partial}=\boldsymbol{\alpha}\cdot\boldsymbol{p}+m\beta$
and $H_{\vec{\boldsymbol{x}}}=-A_{0}+\boldsymbol{\alpha}\cdot\vec{A}$
are implemented in momentum and coordinate space respectively. 

Generally, in one dimension one choose the gauge $A_{0}=A_{0}(z,t)$,
$\vec{A}=\left(0,0,0\right)$ to describe the spatial and temporal
dependent electric field. However, in the case of spatial homogeneous
field, $A_{0}(z,t)$ is linearly depend on space. This also introduce
a spatial dependent variable to the simulation, and bring errors.
So we choose gauge $A_{0}=0$, $\vec{A}=\left(0,0,A_{z}\left(t\right)\right)$
to make $H$ only temporal dependent. In this work, since the field
makes the spin invariant and it is sufficient to focus on the spinless
state in the discussion followed, Dirac matrix in $H$ are reduced
to Pauli matrix. The Hamiltonian is $H=\sigma^{1}\left(p_{z}+A_{z}\left(t\right)\right)+m\sigma^{3}$.
The splitting can be done in two equivalent forms: (a) $H_{\partial}=\sigma^{1}p_{z}+m\sigma^{3}$,
$H_{z}=\sigma^{1}A_{z}(t)$, (b)$H_{\partial}=\sigma^{1}\left(p_{z}+A_{z}\left(t\right)\right)+m\sigma^{3}$,
$H_{z}=0$. Until now, the eigen-states are expressed and the propagating
are done in coordinate space. The pair yield here is an extensive
quantity corresponding to numerical box $L$. To compare with the
QKT results, it should be converted into an intensive quantity, i.e.,
pair yield per Compton wavelength ($\lambda_{C}$). The pair number
should also multiply two for the spin degeneracy.

To reduce the computational cost, in the spatial homogeneous case,
we choose split (b). Then 

\begin{eqnarray}
\exp\left(-idtH_{\partial}\right) & = & \mathcal{F}^{-1}\left\{ \Gamma\left(t\right)\right\} \mathcal{F},\\
\exp\left(-idtH_{z}\right) & = & I,
\end{eqnarray}

\begin{equation}
\Gamma\left(t\right)\equiv\cos\left(\phi\right)-idt\sin\left(\phi\right)\frac{\sigma^{3}m+\sigma^{1}P_{z}\left(t\right)}{\phi},
\end{equation}
where $\phi=dt\sqrt{m^{2}+P_{z}\left(t\right)^{2}}$, $P_{z}\left(t\right)=k_{z}+A\left(t\right)$,
$I$ is unit matrix, and $\mathcal{F}^{-1}(\mathcal{F})$ denotes
(inverse) Fourier transformation. Then the evolution of each temporal
step actually can be done only in momentum space. Using the orthogonality
of the eigen-states, $U_{pn}\left(t\right)$ is diagonal. The pair
number reads $N\left(t\right)=\sum_{k}\left|U_{k}\left(t,t_{0}\right)\right|^{2},$
and

\begin{eqnarray}
U_{k}\left(t,t_{0}\right) & = & \begin{bmatrix}u_{k,a}^{*} & u_{k,b}^{*}\end{bmatrix}\Gamma\left(t\right)...\Gamma\left(t_{0}+dt\right)\Gamma\left(t_{0}\right)\begin{bmatrix}v_{k,a}\\
v_{k,b}
\end{bmatrix},
\end{eqnarray}
where $t_{0}\rightarrow-\infty$, $u_{k,a(b)}$ and $v_{k,a(b)}$
denote the coefficients of the solutions $u_{k}\left(z\right)$ and
$v_{k}\left(z\right)$ of field free Dirac Hamiltonian $H=\sigma^{1}p_{z}+m\sigma^{3}$,
$u_{k}\left(z\right)=e^{ipz}\begin{bmatrix}u_{k,a} & u_{k,b}\end{bmatrix}^{T}$for
positive solutions and $v_{k}\left(z\right)=e^{ikz}\begin{bmatrix}v_{k,a} & v_{k,b}\end{bmatrix}^{T}$
for negative solutions. $u_{k,a}=\sqrt{E_{k}+m}/\sqrt{4\pi E_{k}}$,
$u_{k,b}=sgin\left(k\right)\sqrt{E_{k}-m}/\sqrt{4\pi E_{k}}$, $v_{k,a}=-sgin\left(k\right)\sqrt{E_{k}-m}/\sqrt{4\pi E_{k}}$,
$v_{k,b}=\sqrt{E_{k}+m}/\sqrt{4\pi E_{k}}$, $E_{k}=\sqrt{m^{2}+k^{2}}$.
$\Gamma\left(t\right)$ is a $2\times2$ matrix whose every element
is an function of canonical momentum $k$. Now because the transition
probability $U_{pn}\left(t\right)$ is diagonal, Eq. \eqref{eq:bgTDDE1}
and \eqref{eq:bgTDDE2} degenerate to Eq. \eqref{eq:bgQKT1} and \eqref{eq:bgQKT2}
respectively. This is a convincing evidence of the equivalence between
these two approaches. 

\textbf{Numerical results.}

\begin{figure}
\includegraphics[scale=0.8]{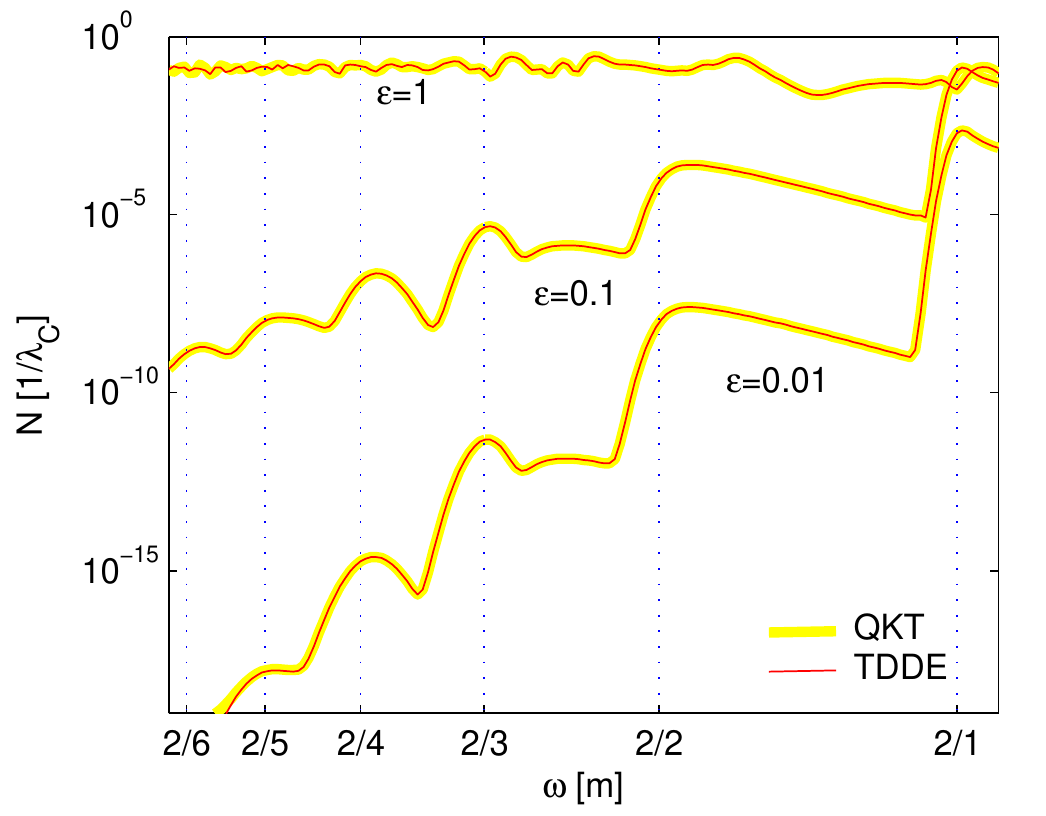}

\caption{Particle yields per Compton wavelength ($\lambda_{C}$) as a function
of frequency $\omega$ computed by QKT and TDDE coincide exactly.
The dashed line indicate the thresholds $2/n$, where $n$ is the
photon number. \label{fig:pairnum}}
\end{figure}

\textbf{\textit{The equivalence.}} We consider a spacial homogeneous
electric pulse of the form $\vec{E}\left(t\right)=\left(0,0,E_{z}\left(t\right)\right)$,

\begin{equation}
E_{z}\left(t\right)=\varepsilon\exp\left(\frac{-t^{2}}{2\tau^{2}}\right)\cos\left(\omega t\right),\label{eq:Et}
\end{equation}
with peak strength $\varepsilon$, duration $\tau$ and frequency
$\omega$. We choose pulse duration $\tau=20/m$ and peak strength
$\varepsilon=0.01,0.1,1m^{2}$. The numerical results are shown in
Fig.\ref{fig:pairnum}. The results have rich physics. For $\varepsilon=0.01,0.1$,
the final pair yield $N\left(t\rightarrow+\infty\right)$ exhibits
a oscillatory structure which is a signature of multiphoton production.
Its thresholds are $n\omega=2m$ (dashed lines in Fig.\ref{fig:pairnum},
$n$ is the photon number, $2m$ is the mass gap). In this log-log
diagram, the linear decay of the pair yield at thresholds as frequency
vanishing indicate a power law decay. Since $N\sim\epsilon^{2n}$,
peaks at $2m/n$ on curve $\varepsilon=0.01$ are $2n$ orders smaller
than on curve $\varepsilon=0.1$. Due to the finite duration $\tau$,
the multiphoton peaks are not sharp. The slightly deviation of the
peaks above $2m/n$ is a signature of the effective mass of the particles
in strong field\citep{effectivemass}. If the field is strong enough,
e.g., $\varepsilon=1$, the mechanism of pair production get in the
tunneling region, and the multiphoton peaks disappear. Of course,
here the collision and back reaction are neglected. As the results
shown, for weak and strong field, namely a large region of Keldysh
parameter, the two approaches coincide exactly. 

\begin{figure}
\includegraphics[scale=0.7]{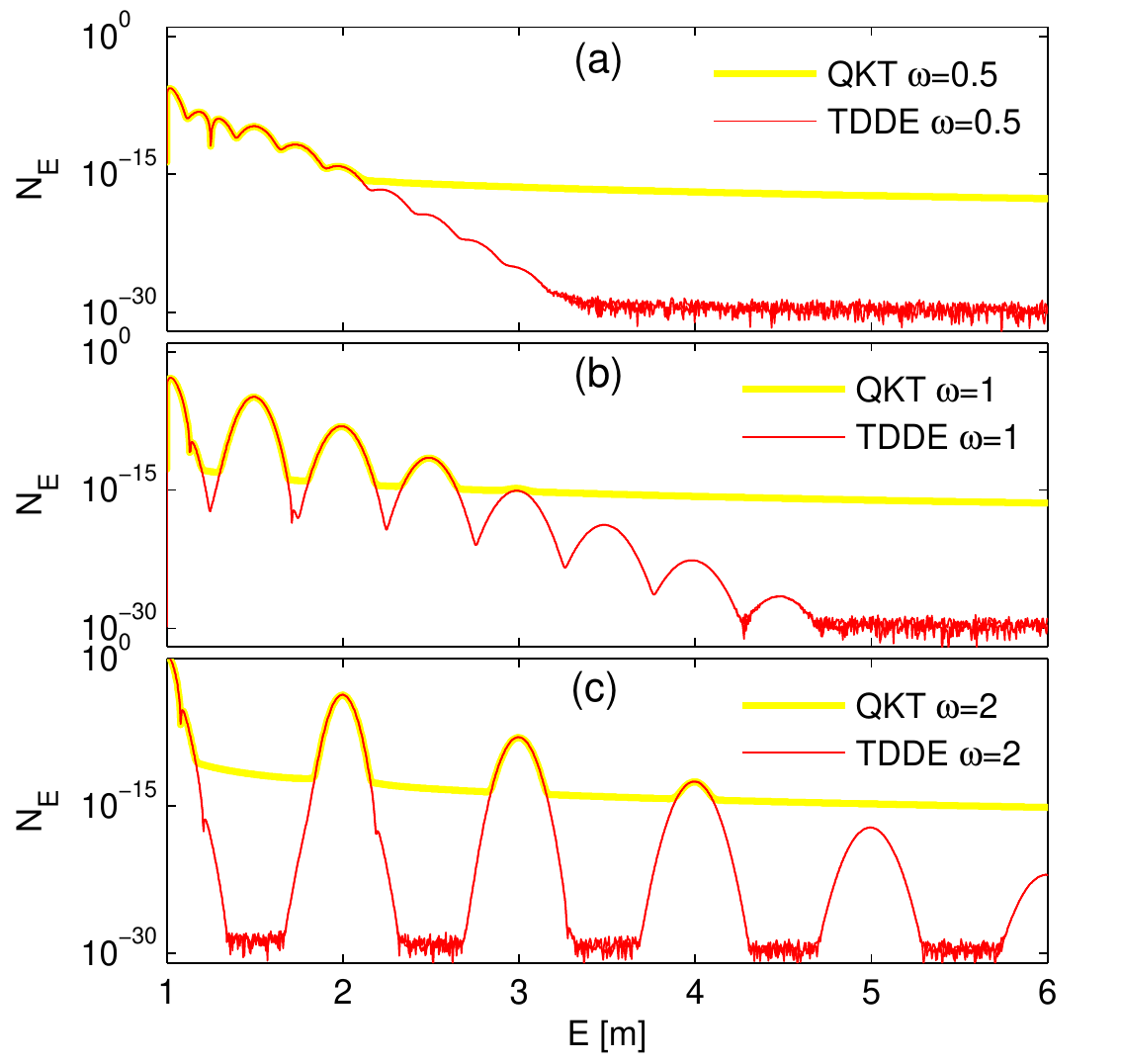}

\caption{The energy distribution for $\epsilon=0.1$, $\omega=0.5,1,2m$. \label{fig:N_E}}
\end{figure}

In order to get more deep insight, we plot the energy distribution
for $\epsilon=0.1$, $\omega=0.5,1,2m$ in Fig.\ref{fig:N_E}, corresponding
to four, two and one photon resonance respectively. In the sub-figure
(a), in addition to the main peak at $E=1$ (zero momentum) which
is due to 4-photon absorption, peaks arise due to the absorption of
$s$ additional photons. Since $(n+s)\omega=2E$ ($E$ is the energy
of single electron, $n=4$, $\omega=0.5m$), the peaks arise at $E=\left(1+s/4\right)m$.
For lower energy, the two approaches coincide exactly, even at the
peak $s=1$ which is split. However, QKT fail to describe higher energy
excitation due to the approximation\citep{EquivalenceS} used in its
derivation. On the other hand, TDDE approach with no approximation
used can provide more accuracy and capture the physics of higher order
multiphoton absorption though it contribute little for the total pair
yield. The high frequency irregular oscillation at the magnitude of
$\sim10^{-30}$ is the limit of precision of TDDE simulation.

\textbf{\textit{Bound states resonance enhanced pair creation.}}\textbf{ }

Now let's turn to the immersed bound states enhanced pair creation.
The spacial homogeneous back ground field takes the form as Eq. \eqref{eq:Et}.
The bound states are supported by a Sauter-like well potential, $V=-A_{0}$,
$\vec{A}=\left(0,0,0\right)$,

\begin{equation}
V\left(z\right)=\frac{V_{0}}{2}\left[\tanh(\frac{z-W/2}{D})-\tanh(\frac{z+W/2}{D})\right],\label{eq:wellpotential}
\end{equation}
where $D$ is the extension of each edge, $W$ is the total width.
We set $D=0.3\lambda_{C}$ and $W=4\lambda_{C}$. For $V_{0}=1m$,
in the well there are bound states of energy $E_{b}=0.19,0.58m$,
and for $V_{0}=0.8m$, $E_{b}=0.37,0.73m$. To avoid time effect\citep{timeeffect},
$V(z)$ is turned on with a modulation coefficient $f(t+T/2)$, $f(t)=\sin^{2}(\pi t/2t_{1})$.
When $t\in[-T/2+t_{1},T/2-t_{1}]$ , $f(t)=1$, that means $V(z)$
is holding on in this time region. Finally, $V(z)$ is turned off
in the way $f(t-(T/2-t_{1}))$, $f(t)=\cos^{2}(\pi t/2t_{1})$. Here
$T$ is total pulse duration. We choose $T=16\tau$ and $t_{1}=\tau$,
$\tau=20/m$ is the pulse duration of Eq. \eqref{eq:Et}. 

\begin{figure}
\includegraphics[scale=0.75]{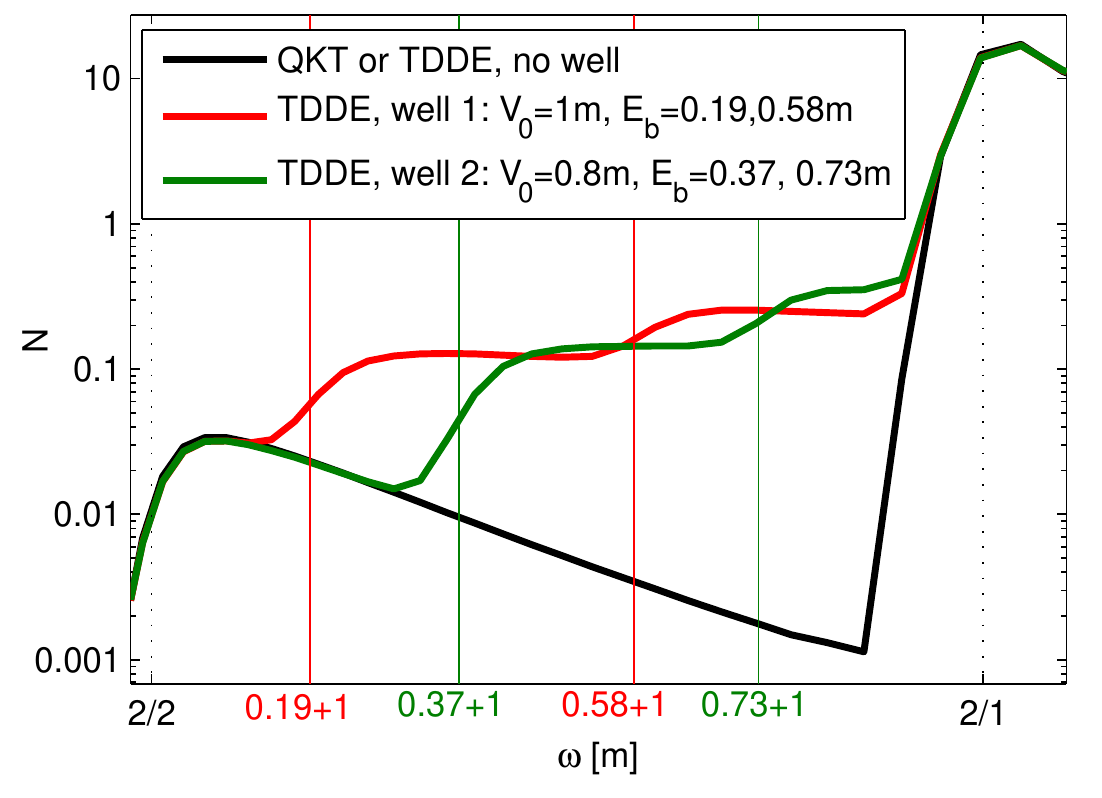}

\caption{Particle yields as a function of frequency $\omega$ computed by TDDE
when the well potential is immersed in the homogeneous back ground
field (red and green lines). The back ground field takes the form
Eq. \eqref{eq:Et} with $\tau=20/m$ , $\varepsilon=0.1m^{2}$, and
its particle yields is also shown here (black line) for comparison.
The well potential takes the form Eq. \eqref{eq:wellpotential} with
$D=0.3\lambda_{C}$ , $W=4\lambda_{C}$ and the turning on and turning
off duration $t_{1}=\tau$. The total duration is $T=16\tau$. The
length of numerical box is $L=137\lambda_{C}$. \label{fig:boundenhacned}}

\end{figure}

If there is only well potential exist, the final pair yield produced
is zero because here it is sub-critical. (Numerically, using TDDE,
the particle yield $<10^{-9}$). The spacial homogeneous electric
pulse, Eq. \eqref{eq:Et} with $\epsilon=0.1m^{2}$, can produce pairs,
as black line in Fig.\ref{fig:boundenhacned}, see also Fig.\ref{fig:pairnum}.
Immersing the well potential in the center of the spacial homogeneous
field of length $L=137\lambda_{C}$, particle yields as a function
of $\omega$ computed by TDDE are shown as red and green lines in
Fig.\ref{fig:boundenhacned}. The vertical fine line indicate $\omega=E_{b}-(-m)$
for the two well respectively . $-m$ is the Dirac sea level. The
results clearly shown that when the photon energy equal to the distance
between bound state and Dirac sea level, pair creation process is
enhanced. The largest enhancement is 2 orders of magnitude for a typical
length scale $L=137\lambda_{C}$. The pair yield can be decomposed
of two parts. One is produced by the background field, and the other
by bound states resonance enhancement. Additionally simulation show
that, at fixed $\omega$ if we increase $L$, the first part increases
linearly, and the second part remains constant. For $V_{0}=1m$, the
max enhancement $\max(N_{enhance})=0.25$ occurs at $\omega=1.7m$,
and for $V_{0}=0.8m$, it is $\max(N_{enhance})=0.35$ at $\omega=1.85m$.
In the region far away from the resonance, the well potential do not
change the pair yield. 

\textbf{Summary.} We have demonstrated that the two widely used approaches
are equivalent. For a homogeneous electric pulse, the particle yield
coincide exactly in a large field strength and frequency region, from
the photon absorption region to the non-perturbative tunneling region,
except that the QKT fails to describe higher energy excitation and
provides less accuracy than TDDE. The details of the approximation
of QKT is left to study in future work. Using TDDE, we studied the
bound states enhanced pair production by immersing bound states into
homogeneous time-dependent background field. For a typical length
scale $L=137\lambda_{C}$, the largest enhancement here is 2 orders
of magnitude, in spite of that two fields are all sub-critical. This
result is helpful for future experiment design. 

Furthermore, due to TDDE's huge computational cost of propagating
all negative energy states in time, only one dimensional system has
been studied until now. In work\citep{QWang}, for spatial dependent
case, we neglect the larger part of the discrete momentum to reduce
computational cost. In this work, for spatial homogeneous case, we
established a formalism which greatly save computing resources. For
examply, in Fig.\ref{fig:N_E}, the simulation of TDDE takes only
2 seconds using matlab on a personal stand-alone computer, even more
faster than the 10 seconds of QKT. This makes it possible that TDDE
can be used to study higher dimensional systems in which more exciting
physics exist. Finally, all discussion in this paper focus on Fermions
described by Dirac equation. The generation to Bosons described by
Klein-Gordon equation is directly.
\begin{acknowledgments}
This work is supported by the National Basic Research Program of China
(Contracts No. 2013CB834100) and the National Natural Science Foundation
of China (Contracts No. 11274051, No. 11374040, No. 11475027, and
No. 11575027).
\end{acknowledgments}

\end{document}